\documentclass[twocolumn,showpacs,floatfix]{revtex4}

\usepackage{graphicx}
\usepackage{bm}
\usepackage{color}
\usepackage{amsmath}
\usepackage{amsfonts}
\usepackage{amssymb}
\usepackage{epstopdf}
\usepackage{ulem}

\begin{document}

\preprint{APS/123-QED}

\title{Experimental Verification of the Universality of Magnetic-Field-Induced Bose\,--\,Einstein Condensation of Magnons} 

\author{Kazuki Shirasawa}
\email{shirasawa@lee.phys.titech.ac.jp}
\author{Nobuyuki Kurita}
\email{kurita.n.aa@m.titech.ac.jp}
\author{Hidekazu Tanaka}
\email{tanaka@lee.phys.titech.ac.jp}
\affiliation{
Department of Physics, Tokyo Institute of Technology, Meguro-ku, Tokyo 152-8551, Japan
}
\date{\today}

\begin{abstract}
CsFeBr$_3$ is an $S\,{=}\,1$ hexagonal antiferromagnet that has a singlet ground state owing to its large easy-plane single-ion anisotropy. The critical behavior of the magnetic-field-induced phase transition for a magnetic field parallel to the $c$ axis, which can be described by the Bose\,--\,Einstein condensation (BEC) of magnons under the $U(1)$ symmetry, was investigated via magnetization and specific heat measurements down to 0.1\,K.
For the specific heat measurement, we have developed a method of effectively suppressing the torque acting on a sample with strong anisotropy that uses the spin dimer compound  Ba$_2$CoSi$_2$O$_6$Cl$_2$ with large and anisotropic Van Vleck paramagnetism.
The temperature dependence of the transition field $H_{\rm c}(T)$ was found to follow the power-law $H_{\rm c}(T)\,{-}\,H_{\rm c}\,{\propto}\,T^{\phi}$ with a critical exponent of ${\phi}\,{=}\,1.50\,{\pm}\,0.02$ and critical field of $H_{\rm c}\,{=}\,2.60$ T.
This result verifies the universality of the three-dimensional BEC of magnons described by ${\phi}_{\rm BEC}\,{=}\,3/2$.
\end{abstract}



\maketitle

\section{Introduction}
Bose\,--\,Einstein condensation (BEC) is a fundamental macroscopic quantum
phenomenon that is a characteristic of systems of bosons~\cite{Bose,Einstein}.
The quantum phase transition (QPT) triggered by applying a magnetic field to a gapped quantum magnet has attracted recent attentions
from the viewpoint of the BEC of magnetic quasiparticles~\cite{Review}.
Gapped quantum magnets that have been reported to exhibit field-induced QPTs generally have an excitation gap $\Delta$ between a disordered singlet ground state and degenerate lowest excited states. The origins of the gap are the dimerization of a spin pair by a strong antiferromagnetic exchange interaction, and the large easy-plane single-ion anisotropy $D(S^z)^2$ for systems with $S\geq 1$.
The application of a magnetic field lifts the degeneracy of the excited states by the Zeeman effect and decreases the magnitude of $\Delta$. 
At the critical field $H_{\rm c}$ ($\,={\Delta}/g\mu_{\rm B}$), where the gap closes completely at $T=0$, an antiferromagnetic (AF) ordering occurs in a plane perpendicular to the magnetic field owing to the coherent superposition of the singlet and magnetic component of the lowest excited state.
Because the excited magnetic components, magnons, can be treated as interacting lattice bosons\,\cite{Rice}, the magnetic-field-induced QPT from the disordered singlet state to the transverse AF state can be described by the BEC of magnons\,\cite{Affleck_PRB1991,Giamarchi_PRB1999,Nikuni_PRL2000,Ruegg_Nature2003}. 
In magnon BEC, the particle quantities of the chemical potential $\mu$ and boson density $n$ correspond to the magnetic quantities of the external magnetic field $H$ and total magnetization $M$, respectively.


Experimentally, magnon BEC can be identified by distinct features beyond the framework of the conventional mean field model\,\cite{TY,TM}, such as a cusplike minimum in the magnetization at the transition temperature ($T_{\rm N}$) and power-law behavior of the low-temperature phase boundary in the magnetic field vs temperature ($H\,{-}\,T$) phase diagram\,\cite{Oosawa_JPCM1999}.
Regarding the latter, the phase boundary satisfies the formula 
\begin{equation}
H_{\rm c}(T)\,{-}\,H_{\rm c}\,{\propto}\,T^{\phi}, 
\label{power}
\end{equation}
where $H_{\rm c}(T)$ is the transition field at temperature $T$.
This expression, which assumes a dilute boson limit, is valid only at sufficiently low temperatures compared with the energy scale of AF exchange couplings or boson interactions.
It has been theoretically predicted that the critical exponent $\phi$ is given by $\phi_{\rm BEC}\,{=}\,3/2$ for magnon BEC in a three-dimensional (3D) system\,\cite{Affleck_PRB1991,Giamarchi_PRB1999,Nikuni_PRL2000}.
Thus far, however, the precise experimental study of the universality of 3D magnon BEC has been limited to a few compounds including the $S\,{=}\,1/2$ interacting spin dimer system TlCuCl$_3$\,~\cite{YamadaF_JPSJ2008,YamadaF2}, the $S$\,=\,1 spin chain system NiCl$_2$-4SC(NH$_2$)$_2$ (abbreviated to DTN) with large uniaxial easy-plane single-ion anisotropy~\cite{Zapf_PRL2006,Yin_PRL2008,Blinder_PRBR2017}, and the $S\,{=}\,1/2$ four-leg spin ladder system Cu$_2$Cl$_4$-H$_8$C$_4$SO$_2$ (abbreviated to Sul-Cu$_2$Cl$_4$)\,\cite{Fujisawa_JPSJ2003,Fujisawa_PTP2005,Povarov_PRBR2015}.
This is mainly due to the technical difficulty in reaching the required low temperatures and in performing precise measurements under high magnetic fields exceeding $H_{\rm c}$.
In fact, when the analyzed range of $T$ is not sufficiently low, the $\phi$ value can be overestimated\,~\cite{Nohadani_PRB2004,Nohadani_PRB2005,Kawashima_JPSJ2004,Kawashima_JPSJS2005}, as in earlier reports on TlCuCl$_3$\,\cite{Oosawa_JPCM1999,Oosawa_PRB2001,TanakaH_JPSJ2001,Shindo_JPSJ2004} and DTN\,\cite{Paduan-Filho_JAP2004,Paduan-Filho_PRB2004}.

In the strict sense, magnon BEC requires uniaxial symmetry around the applied magnetic field, $O(2)$, which corresponds to the conservation of the total number of particles, $U(1)$.
However, the $O(2)$ symmetry is not necessarily satisfied even in TlCuCl$_3$ and DTN because of their low-symmetry crystal lattice, magneto-elastic coupling above $H_{\rm c}$ caused by a collinear spin ordering perpendicular to the symmetric axis, and lattice disorder which is considered to be a factor contributing to the controversial results on the BEC universality of DTN\,\cite{Yin_PRL2008,Wulf}.
As discussed in Ref.\,\onlinecite{Dell'Amore_PRB2009}, the magnetic-elastic effect unavoidably present in any compound results in a weakly first-order-like behavior at the magnetic phase transition. 
On this point, Sul-Cu$_2$Cl$_4$ is suggested to be free from the above mentioned anisotropy or magneto-elastic effect with the help of its unique electric properties\,\cite{Povarov_PRBR2015}.

The singlet-ground-state antiferromagnet CsFeBr$_3$ with the hexagonal crystal structure~\cite{Takeda_JPSJ1974} provides an excellent opportunity for investigating magnon BEC.
In this compound, magnetic Fe$^{2+}$ ions surrounded by six Br$^-$ ions form AF chains along the $c$ axis, and a uniform triangular lattice in the $ab$ plane with AF exchange interactions.
Owing to the strong spin-orbit coupling and trigonal crystal distortion,
the effective magnetic moment of Fe$^{2+}$ ions can be described by the pseudospin $S$\,=\,1
at low temperatures considerably below $|\lambda|/k_{\rm B}$\,$\approx$\,150\,K, where $\lambda$ represents the spin-orbit coupling constant\,~\cite{Inomata_JPSJ1967}.
The Hamiltonian is given by
\begin{eqnarray}
{\cal H}&=&\sum_i D\left(S_i^z\right)^2+\sum_{\langle i,j\rangle}^{\rm chain} J_0\left(S_i^xS_j^x+S_i^yS_j^y+{\Delta}S_i^zS_j^z\right)\nonumber\\
&+&\sum_{\langle l,m\rangle}^{\rm plane} J_1\left(S_l^xS_m^x+S_l^yS_m^y+{\Delta}S_l^zS_m^z\right),
\end{eqnarray}
where $D$ ($>$\,0) corresponds to the energy difference between the singlet ground state $|S^z$=0$\rangle$ and the doublet first excited states $|S^z$=$\pm 1\rangle$, and $J_0$ ($>$\,0) and $J_1$ ($>$\,0) represent the AF exchange interactions in the chain and $ab$ plane, respectively. Here, the $z$ axis is taken to be parallel to the crystallographic $c$ axis. The magnitude of the $D$-term is considerably larger than that of the exchange interactions, which leads to a gapped ground state\,~\cite{Bocquet_PRB1988,Dorner_ZPhysB1988}. CsFeBr$_3$ undergoes a magnetic ordering corresponding to magnon BEC when subjected to an external magnetic field parallel to the $c$ axis. The spin configuration in the ordered state above $H_{\rm c}\,{\sim}\,3$ T~\cite{TanakaY_JPSJ2001} is the canted 120$^\circ$ structure characteristic of a triangular lattice antiferromagnet~\cite{Dorner_ZPhysB1990}.

In CsFeBr$_3$, the in-plane anisotropy, which breaks the $O(2)$ symmetry, is expected to be negligibly small, even in the ordered phase above $H_{\rm c}$, because the spin configuration has hexagonal symmetry~\cite{Dorner_ZPhysB1990}. Thus, the hexagonal symmetry of the magnetic interactions is maintained across $H_{\rm c}$, even if magnetic elastic coupling exists. In addition, its $H_{\rm c}$ value of less than 3\,T makes it possible to carry out a variety of experiments using commercial superconducting magnets. For these reasons, CsFeBr$_3$ is advantageous for precise verification of the universality of 3D magnon BEC. Although CsFeBr$_3$ has long been known to exhibit a magnetic-field-induced AF ordering~\cite{Dorner_ZPhysB1990}, little attentions has been paid to its low-temperature quantum critical phenomena except in our previous report~\cite{TanakaY_JPSJ2001}. 

In this paper, we have determined the $H-T$ phase diagram of CsFeBr$_3$ via low-temperature magnetic and thermal measurements down to 0.1\,K and have investigated the quantum critical phenomenon for field-induced AF ordering.
The low-temperature phase boundary was found to satisfy the power law expressed by Eq.\,(\ref{power}).
The exponent $\phi$ obtained by fitting to the data unambiguously converges to ${\phi}\,{=}\,1.50\,{\pm}\,0.02$, 
which is in excellent agreement with the universality of 3D magnon BEC.
We also report a novel method of measuring the high-field specific heat of magnetic compounds with large magnetic anisotropy.

\section{Experimental details}
Single crystals of CsFeBr$_3$ were grown by the vertical Bridgman method from a melt of a stoichiometric mixture of the constituent elements sealed in an evacuated quartz
tube\,\cite{TanakaY_JPSJ2001}. 
The quartz tube was lowered in the vertical direction at a rate of 3\,mm/h from the center of the furnace, the temperature of which was regulated at 540$^\circ$C.
After the removal of impurities and imperfect crystals, we repeated the same procedure.

The specific heat $C$ was measured at temperatures down to 0.1\,K in magnetic fields of up to 9\,T using a physical property measurement system (PPMS, Quantum Design) by a relaxation method. 
As will be explained below, the spin dimer compound Ba$_2$CoSi$_2$O$_6$Cl$_2$ (BCSOC)\,\cite{Tanaka_JPSJ2014} was employed to cancel out the large magnetic torque originating from the CsFeBr$_3$ single crystal.
The background contributions were subtracted.
The magnetization $M(T, H)$ was measured down to $T$\,=\,0.5\,K and up to $\mu_0H$\,=\,7\,T 
using a SQUID magnetometer (MPMS-XL, Quantum Design) equipped with a $^3$He device (iHelium3, IQUANTUM). 
Magnetic fields were applied parallel to the $c$ axis of CsFeBr$_3$ for both measurements.

\section{Results and Discussion}
In a previous specific heat study by a relaxation method\,\cite{TanakaY_JPSJ2001}, 
a two-step anomaly was observed
in high-fields for $H$\,$\parallel$\,$c$.
This implied the presence of an exotic intermediate state in CsFeBr$_3$.
After the measurement, however, we noticed that some of the thin metal wires which loosely hook a sample platform were disconnected or had loosened.
This indicates that the platform moved or rotated during measurements under high fields, probably due to the large magnetic torque acting on CsFeBr$_3$, which has large anisotropy in its magnetic susceptibility, ${\chi}_{ab}\,{\gg}\,{\chi}_{c}$.
We concluded that in the previous specific heat measurement\,\cite{TanakaY_JPSJ2001}, the crystallographic $c$ axis was inclined from the magnetic field, which led to the two-step phase transition. This is also supported by the observation of a single magnetization anomaly\,\cite{TanakaY_JPSJ2001}.

In this study, we employed plate-shaped single crystals of BCSOC to suppress the movement of CsFeBr$_3$ under high magnetic fields. BCSOC, which has a layered structure, has a nonmagnetic gapped ground state, which remains under magnetic fields up to a critical field of 32\,T for $H$\,$\parallel$\,$ab$ and 47\,T for $H$\,$\parallel$\,$c^*$. BCSOC crystals are easily cleaved parallel to the $ab$ plane.
The Van Vleck paramagnetism of BCSOC is large and highly anisotropic with $\chi_{\rm VV}$=1$\times 10^{-2}$ and 1$\times 10^{-3}$\,emu/mol for $H$\,$\parallel$\,$ab$ and $H$\,$\parallel$\,$c^*$, respectively\,\cite{Tanaka_JPSJ2014}. Therefore, the magnetic torque makes the $ab$ plane of BCSOC parallel to the field direction.

\begin{figure}
\begin{center}
\includegraphics[width=1\linewidth]{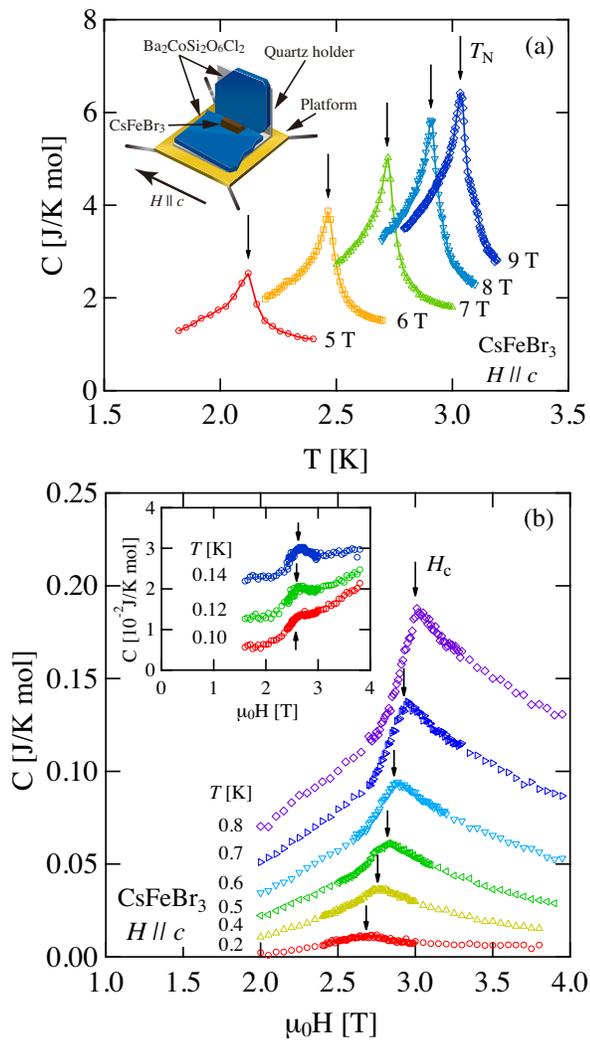}
\end{center}
\caption{(Color online) 
(a) $C$ vs $T$ for CsFeBr$_3$ under several magnetic fields with $H \parallel c$.
The inset illustrates the experimental setup for the specific heat measurement.
Two plate-shaped Ba$_2$CoSi$_2$O$_6$Cl$_2$ single crystals were used to suppress the movement of the CsFeBr$_3$ single crystal under high fields. See text for details.
(b) $C$ vs $\mu_0 H$ for CsFeBr$_3$ with $H \parallel c$ at several fixed temperatures.
The inset shows $C(H)$ data at lower temperatures down to 0.10\,K where the $C(H)$ data are arbitrarily shifted upward for clarity.
The vertical arrows in (a) and (b) indicate the transition temperature $T_{\rm N}(H)$ and field $H_{\rm c}(T)$, respectively.
} \label{fig1}
\end{figure}

The inset of Fig.~\ref{fig1}(a) illustrates the experimental setup for the present specific heat measurements.
Two BCSOC plates were arranged perpendicularly to each other using an L-shaped quartz plate placed on a sample stage. A magnetic field was applied parallel to both plate faces. In this configuration, the magnetic torque acting on the two BCSOC plates makes their intersection side parallel to the field direction. A small piece of CsFeBr$_3$ single crystal of $\sim$\,0.5\,mg was placed on a BCSOC plate with its $c$ axis parallel to the intersection.
Apiezon N grease was used to ensure both mechanical and thermal contact.
A single BCSOC plate with mass and dimensions of approximately 10\,mg and $3\times 3 \times 0.5\,$mm$^3$, respectively, was confirmed to be sufficient to suppress the movement of CsFeBr$_3$ under magnetic fields for $H$\,$\parallel$\,$c$.
We also applied this technique to specific heat measurements of the related compounds CsFeCl$_3$ and RbFeBr$_3$, and confirmed that the obtained $H\,{-}\,T$ phase diagrams are consistent with those determined from magnetization data\,\cite{Kurita_unpublished}.

The main panel of Fig.~\ref{fig1}(a) shows the $T$ dependence of the specific heat $C(T)$ measured under several magnetic fields for $H$\,$\parallel$\,$c$.
In contrast to the previous results\,\cite{TanakaY_JPSJ2001}, only a single peak at $T_{\rm N}$ was identified in $C(T)$ under magnetic fields of up to 9\,T. 
As can be seen in Fig.~\ref{fig1}(b), the $H$ dependence of the specific heat $C(H)$ also shows a single peak at a transition field $H_{\rm c}(T)$ down to 0.1\,K.
Thus, we conclude that the magnetic-field-induced ordered phase of CsFeBr$_3$ constitutes a single AF phase when the magnetic field is precisely parallel to the $c$ axis.
We infer that owing to the $c$-plane component of the magnetic field $H_{\perp}$ produced by the inclination of the crystal, the spin components perpendicular to $H_{\perp}$ are first ordered with decreasing temperature, and then the spin component parallel to $H_{\perp}$ is ordered, which leads to the two-step ordering observed in the previous measurement\,\cite{TanakaY_JPSJ2001}.

\begin{figure}
\begin{center}
\includegraphics[width=1\linewidth]{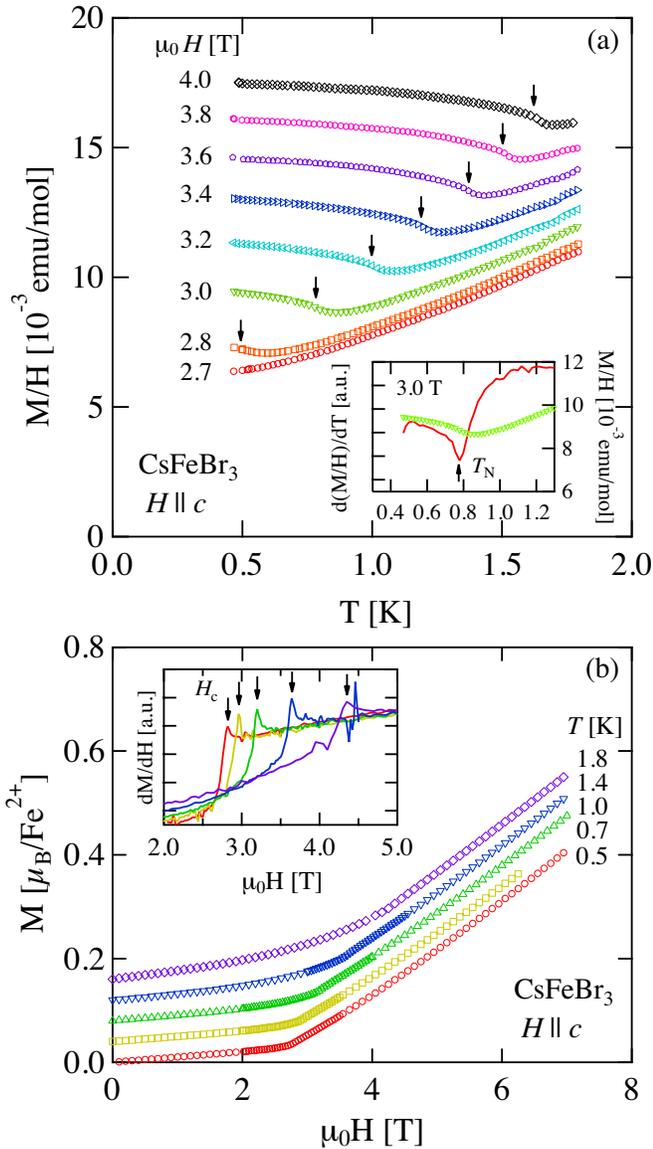}
\end{center}
\caption{(Color online) 
(a) Temperature dependence of $M/H$ for CsFeBr$_3$ under several magnetic fields with $H \parallel c$.
As demonstrated in the inset using the 3.0\,T data, $T_{\rm N}$ is assigned to the temperature giving the peak in d($M/H$)/d$T$($T$).
(b) Magnetization curves of CsFeBr$_3$ for $H \parallel c$ at several fixed temperatures. The data are shifted upward by multiples of $4\,{\times}\,10^{-2}$\,${\mu}_{\rm B}$/Fe$^{2+}$.
The inset shows d$M/$d$H(H)$ vs $H$ at selected temperatures.
Vertical arrows indicate the transition field $H_{\rm c}(T)$.
} \label{fig2}
\end{figure}

Figure~\ref{fig2}(a) shows the $T$ dependence of the magnetic susceptibility $\chi$ $(=M/H)$ of CsFeBr$_3$ under several fields for $H$\,$\parallel$\,$c$. 
$\chi(T)$ monotonically decreases down to the lowest $T$ of 0.5\,K under a magnetic field of 2.7\,T, indicating that the ground state remains a gapped singlet up to 2.7\,T.
For higher magnetic fields of above 2.7\,T, $\chi(T)$ shows a cusplike minimum indicative of an AF ordering. The cusplike minimum in $\chi(T)$ is a characteristic of magnon BEC\,~\cite{Nikuni_PRL2000} and is commonly observed in gapped quantum magnets exhibiting magnon BEC\,~\cite{Oosawa_JPCM1999,Paduan-Filho_PRB2004,Waki_JPSJ2004,Oosawa_PRB2002,Kurita_PRB2016}.
As displayed in the inset of Fig.~\ref{fig2} using the 3.0\,T data, we assign $T_{\rm N}$ to the temperature giving the peak in d$\chi$/d$T(T)$, because this temperature coincides with the temperature giving a peak in the specific heat. The transition temperature $T_{\rm N}$ increases with increasing field as indicated by arrows.

Figure~\ref{fig2}(b) shows the magnetization curves $M(H)$ measured at several fixed $T$ down to 0.5\,K.
At 0.5\,K, a phase transition from the gapped state to the AF state appears as a clear kink-like anomaly in $M(H)$ at approximately 2.7\,T.
The transition field $H_{\rm c}(T)$ is defined as the field giving the peak in d$M/$d$H(H)$ as shown in the inset of Fig.~\ref{fig2}(b).
With increasing temperature, $H_{\rm c}(T)$ shifts to a higher field, while the peak anomaly becomes smeared.
The finite slope of $M(H)$ in the gapped state below $H_{\rm c}(T)$ arises from the large Van Vleck paramagnetism, as observed in the related compound CsFeCl$_3$\,\cite{Kurita_PRB2016}.

\begin{figure}
\begin{center}
\includegraphics[width=1\linewidth]{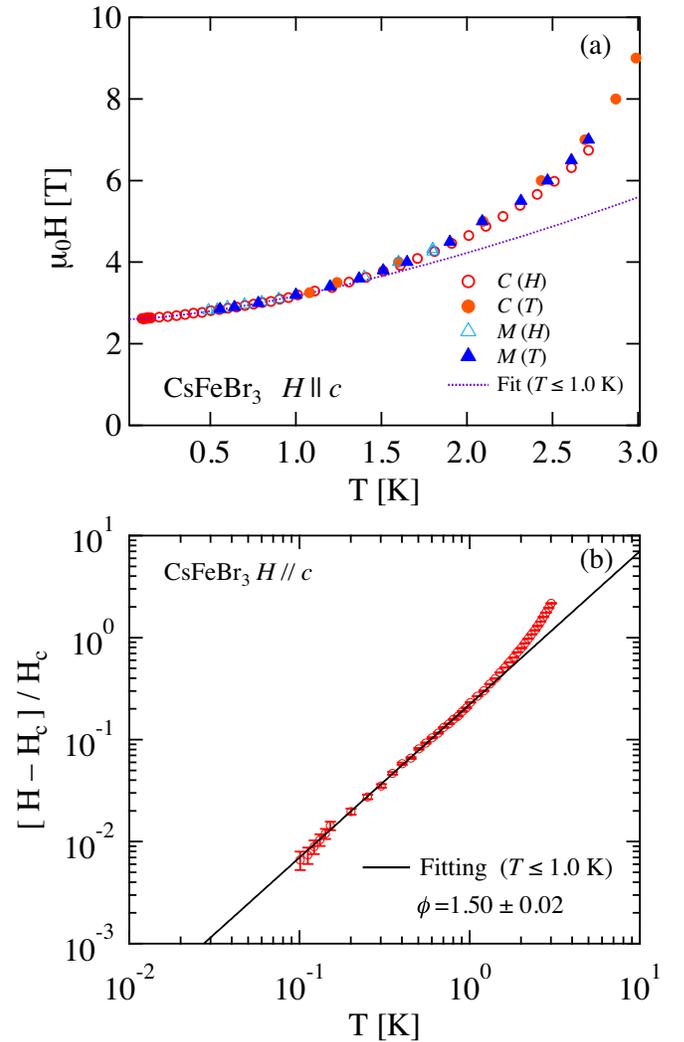}
\end{center}
\caption{(Color online)
(a) Magnetic field vs temperature phase diagram of CsFeBr$_3$ for $H$\,$\parallel$\,$c$ 
determined from specific heat $C(T,H)$ and magnetization $M(T,H)$ measurements.
The dotted curve represents a fit to the power law in Eq.~(\ref{power}) with $\phi$\,=\,1.50 using data below 1.0\,K.
(b) The reduced field $(H-H_{\rm c})$/$H_{\rm c}$ vs $T$ on a double logarithmic scale.
The solid line is a fit with $\phi$\,=\,1.50.
} \label{fig3}
\end{figure}

The $H$\,$-$\,$T$ phase diagram of CsFeBr$_3$ with $H$\,$\parallel$\,$c$ was obtained as shown in Fig.~\ref{fig3}(a).
The transition points $T_{\rm N}(H)$ and $H_{\rm c}(T)$ determined by two different measurements are in good agreement.
It was found that the low-temperature phase boundary follows the power law expressed by Eq.~(\ref{power}).
The dotted curve is a fit to the data with $\phi$\,=\,1.50 and $H_{\rm c}$\,=\,2.60\,T.
To determine the critical exponent $\phi$ for $T$\,$\rightarrow$\,0, 
the $\phi$ value was evaluated from a best fit with the power law in various $T$ regimes between $T_{\rm min}$ fixed to 0.1\,K and $T_{\rm max}$ ranging from 0.5 to 2.0\,K.
In this analysis, $H_{\rm c}$ was used as a free parameter.
With decreasing $T_{\rm max}$, $\phi$ and $H_{\rm c}$ monotonically decrease and 
converge to 1.50\,$\pm$\,0.02 and 2.60\,T, respectively, below 1\,K. 
Note that below 1\,K, the error for $\phi$ increases with decreasing $T_{\rm max}$ owing to the insufficient number of data points. 
We also performed power-law analysis while fixing $H_{\rm c}$ close to 2.60\,T. As shown in Fig.~\ref{fig3}(b), the power-law behavior can be more clearly observed in the double-logarithmic plot of the scaled field $(H-H_{\rm c})$/$H_{\rm c}$ against $T$.
From these analyses, we confirmed that $\phi$ converges to 1.50\,$\pm$\,0.02 with $H_{\rm c}$\,=\,2.60\,T as $T$\,$\rightarrow$\,0, which is in excellent agreement with the universality of 3D magnon BEC described by ${\phi}_{\rm BEC}\,{=}\,1.5$\,~\cite{Affleck_PRB1991,Giamarchi_PRB1999,Nikuni_PRL2000}.


\section{Summary}
We have determined the $H$\,--\,$T$ phase diagram of the CsFeBr$_3$ with a gapped singlet ground state by low-temperature magnetization and specific heat measurements down to 0.1\,K. To suppress the rotation of the sample induced by magnetic torque in the specific measurement, we used two plate-shaped single crystals of the dimerized quantum magnet Ba$_2$CoSi$_2$O$_6$Cl$_2$, which has large and anisotropic Van Vleck paramagnetism. Consequently, we found that, contrary to in the previous report, CsFeBr$_3$ exhibits only a single phase transition for $H\,{\parallel}\,c$. Because the hexagonal symmetry of the crystal lattice of CsFeBr$_3$ is maintained even in the ordered phase above the critical field $H_{\rm c}$, the anisotropy that breaks the $U(1)$ symmetry required for magnon BEC will be negligibly small. Power-law analysis of the lower-temperature phase boundary using Eq.~(\ref{power}) showed that the exponent $\phi$ converges unambiguously to 1.50$\pm$0.02 as $T$\,$\rightarrow$\,0, which is in excellent agreement with the critical exponent ${\phi}_{\rm BEC}\,{=}\,3/2$ for 3D magnon BEC. This result verifies the universality of 3D magnon BEC.

\section*{Acknowledgments}
We thank R. Takeda for his technical support. This work was supported by Grants-in-Aid for Scientific Research (A) (Nos. 23244072 and 26247058) and (C) (No. 16K05414) and a Grant-in-Aid for Young Scientists (B) (No. 26800181) from Japan Society for the Promotion of Science.


\end{document}